\documentclass{aa}  

\usepackage{graphicx,natbib}
\bibpunct{(}{)}{;}{a}{}{,} 
\usepackage[varg]{txfonts}
\begin{document}

   \title{Evidence of a primordial isotopic gradient in the inner region of the solar protoplanetary disc}

   \author{J. Mah\inst{1,2}
          \and
          R. Brasser\inst{3}
          \and
          J.M.Y. Woo\inst{4,5}
          \and
          A. Bouvier\inst{6}
          \and
          S.J. Mojzsis\inst{3,7,8}
          }

   \institute{Max-Planck-Institut f\"{u}r Astronomie, K\"{o}nigstuhl 17, 69117 Heidelberg, Germany\\
             \email{mah@mpia.de}
             \and
             Earth Life Science Institute, Tokyo Institute of Technology, Tokyo 152-8550, Japan
             \and
             Origins Research Institute, Research Centre for Astronomy and Earth Sciences, 15-17 Konkoly Mikl\'{o}s Thege utca, 1121 Budapest, Hungary
             \and
             Institut f\"{u}r Planetologie, University of M\"{u}nster, Wilhelm-Klemm-Str. 10, 48149 M\"{u}nster, Germany
             \and
             Laboratoire Lagrange, Universit\'{e} C\^{o}te d'Azur, CNRS, Observatoire de la C\^{o}te d'Azur, 06304 Nice, France
             \and
             Bayerisches Geoinstitut, Universit\"{a}t Bayreuth, 95447 Bayreuth, Germany
             \and
              Department of Lithospheric Research, University of Vienna, 1090 Vienna, Austria
             \and
              Department of Geological Sciences, University of Colorado Boulder, Boulder, CO 80309-0399, USA     
             }

   \date{Received 16 December 2021; 22 February 2022}

 
  \abstract
  {Not only do the sampled terrestrial worlds (Earth, Mars, and asteroid 4 Vesta) differ in their mass-independent (nucleosynthetic) isotopic compositions of many elements (e.g. $\varepsilon$\element[][48]{Ca}, $\varepsilon$\element[][50]{Ti}, $\varepsilon$\element[][54]{Cr}, $\varepsilon$\element[][92]{Mo}), the magnitudes of some of these isotopic anomalies also appear to correlate with heliocentric distance. While the isotopic differences between the Earth and Mars may be readily accounted for by the accretion of mostly local materials in distinct regions of the protoplanetary disc, it is unclear whether this also applies to asteroid Vesta. Here we analysed the available data from our numerical simulation database to determine the formation location of Vesta in the framework of three planet-formation models: classical, Grand Tack, and Depleted Disc. We find that Vesta has a high probability of forming locally in the asteroid belt in models where material mixing in the inner disc is limited; this limited mixing is implied by the isotopic differences between the Earth and Mars. Based on our results, we propose several criteria to explain the apparent correlation between the different nucleosynthetic isotopic compositions of the Earth, Mars, and Vesta: (1) these planetary bodies accreted their building blocks in different regions of the disc, (2) the inner disc is characterised by an isotopic gradient, and (3) the isotopic gradient was preserved during the formation of these planetary bodies and was not diluted by material mixing in the disc (e.g. via giant planet migration).}


   \titlerunning{Evidence for incipient isotopic gradient in inner disc}

   \maketitle
%

\section{Introduction}
The bulk isotopic and elemental compositions of a planetary body are the cumulative average of its building blocks \citep{DrakeRighter2002,Fitoussietal2016,Dauphas2017,Mezgeretal2020}, which in turn is the end-product of the protoplanetary disc region(s) from which it accreted. The sampled Solar System objects (i.e. rocks and meteorites) for which we know (or infer) the parent bodies are: Earth, Moon, Mars, and asteroid 4 Vesta. Furthermore, asteroids 434 Hungaria and 6 Hebe may also be represented in our meteorite collections \citep{Greenwoodetal2020} because they are thought to be the parent bodies of the aubrites \citep{Zellner1975,Zellneretal1977,Clarketal2004,Cuketal2014} and H chondrites \citep[one of the components of the ordinary chondrite group;][]{Gaffeyetal1993,GaffeyGilbert1998,Binzeletal2004,Binzeletal2019}, respectively. Together, the meteorites representing these planetary bodies fall into a larger group known as the non-carbonaceous (NC), or terrestrial, group based on their bulk isotopic compositions \citep{Warren2011}. The constituents of the NC group are thought to have originated from the inner Solar System, whereas the carbonaceous (C), or Jovian, group represents the outer Solar System \citep{BrasserMojzsis2020}.

Although the Earth, Mars, the Howardite–Eucrite–Diogenite (HED) clan of meteorites \citep[likely tied to asteroid 4 Vesta;][]{McCordetal1970,ConsolmagnoDrake1977,BinzelXu1993,Keil2002,McSweenetal2013}, the aubrites, and the H chondrites are broadly part of the NC group, their isotopic compositions are not identical. There are measurable differences in their mass-independent isotope anomalies on the scale of parts per $10^4$ to $10^6$ for Ba, Ca, Cr, Fe, Mo, N, Nd, Ni, O, Ru, Sm, Ti, W, Xe, and Zr \citep[see][for a review]{QinCarlson2016}. The variations in nucleosynthetic contributions for many elements are due to the particular distribution of presolar dust grains and/or irradiation processes (e.g. for the elements O and Ti) in the solar protoplanetary disc; they are considered to be impervious to planetary processes because they are mass-independently fractionated in their isotope ratios, which means that they can only be lost by dilution (i.e. via redistribution of material in the disc). Since we observe nucleosynthetic anomalies for samples from the Earth, Mars, the HEDs, the aubrites, and the H chondrites across a range of lithophile and siderophile elements, it suggests that the origins of these anomalies are very likely to be primordial. Furthermore, two planetary bodies will only exhibit the same isotopic composition if they both accreted their building blocks from the same reservoir(s) with the same proportion of nucleosynthetic isotopes. Therefore, the observed differences in the isotopic compositions of Earth, Mars, and asteroid Vesta (and possibly asteroids Hungaria and Hebe) suggest that there are differences in the compositions of their building blocks that would imply formation at various locations throughout the disc \citep[e.g.][]{Carlsonetal2018,Mezgeretal2020}. It is worth noting, however, that at the present time no cosmochemical model can simultaneously account for both the bulk and isotopic compositions of the planets and the different meteorite groups by any mixing model of the known components.

\begin{figure*}
\centering
   \resizebox{0.9\hsize}{!}{\includegraphics{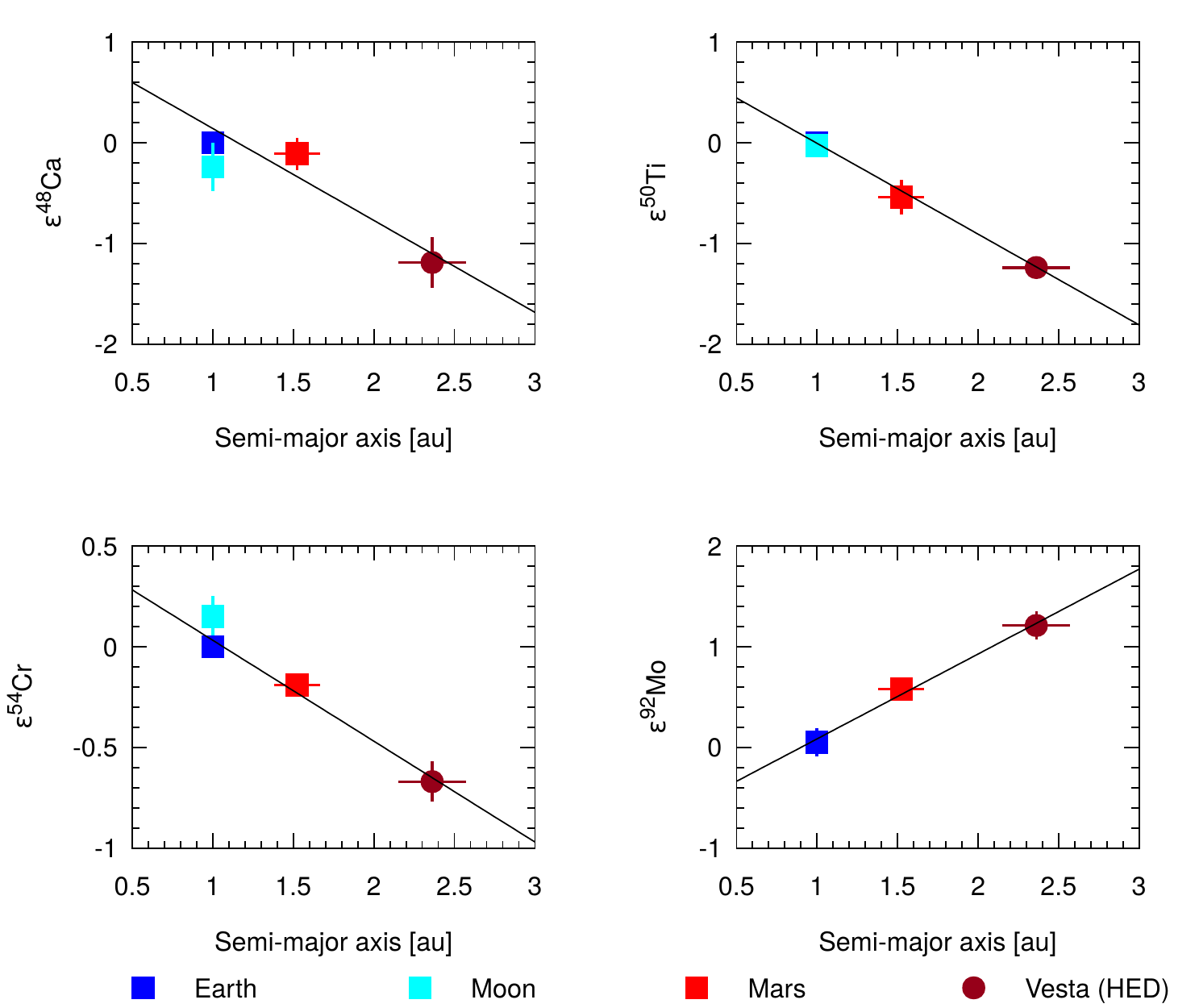}}
   \caption{Calcium (\element[][48]{Ca}), titanium (\element[][50]{Ti}), chromium (\element[][54]{Cr}), and molybdenum (\element[][92]{Mo}) isotopic anomalies for Earth, Moon, Mars, and Vesta (sampled by the HED meteorites) plotted with respect to their distance from the Sun. The isotopic compositions of these planetary bodies appear to correlate with their semi-major axes, alluding to the presence of isotopic gradients in the inner Solar System \citep[e.g.][]{Trinquieretal2009,Yamakawaetal2010}; the data are from the compilation of Dauphas (2017) and references therein. The $\varepsilon$\element[][92]{Mo} value for Mars is from \citet{Burkhardtetal2021}, while the value for HED is plotted using data for Mesosiderites sourced from \citet{Dauphasetal2002} because there are no known Mo measurements for HED meteorites to date. There are also no known Mo measurements for the Moon.}
   \label{fig:correlation}
\end{figure*}

Several prior works \citep[e.g.][]{Trinquieretal2009,Yamakawaetal2010} have identified an apparent correlation between the isotopic anomalies in $\varepsilon$\element[][50]{Ti} and $\varepsilon$\element[][54]{Cr} and the semi-major axis of the Earth, Mars, and Vesta (the HED meteorites). The reported correlation also exists for $\varepsilon$\element[][48]{Ca} \citep[e.g.][]{Schilleretal2018} and possibly for $\varepsilon$\element[][92]{Mo} \citep[e.g.][]{Burkhardtetal2011}, as shown in Fig.~\ref{fig:correlation}. The origin of the observed correlations in \element[][50]{Ti} and \element[][54]{Cr} favoured by \citet{Yamakawaetal2010} is that Type 1a supernovae supplied these isotopes to the inner Solar System just before the formation of the planetesimals that became the precursors of Earth, Mars, and Vesta, although the supernova origin view is debated \citep{Wasserburgetal2006}. \citet{Yamakawaetal2010} argued for a short timing of isotope delivery due to the rapid homogenisation of nuclides in the protoplanetary disc revealed by results of hydrodynamical simulations: radionuclides injected by supernovae spread out evenly in the protoplanetary disc on timescales of $10^3$ to $10^6$ years \citep[e.g.][and references therein]{Ouelletteetal2009}. The correlated isotopic compositions of the Earth, Mars, and Vesta therefore reflect a potential gradient in the distribution of \element[][48]{Ca}, \element[][50]{Ti}, and \element[][54]{Cr} isotopes in the inner protoplanetary disc where there is an increased depletion of these isotopes with increasing distance from the Sun. The trend for $\varepsilon$\element[][92]{Mo} in Fig.~\ref{fig:correlation} is different and could be due to the fact that Mo isotopes are synthesised via the $p$-, $s$-, and $r$-processes \citep[e.g.][]{Brenneckaetal2013}, in contrast to the suggested supernova origin for the \element[][48]{Ca}, \element[][50]{Ti}, and \element[][54]{Cr} isotopes \citep[e.g.][]{Hartmannetal1985,Woosley1997,Wanajoetal2013}.

The origin of the observed diversity in the isotopic anomalies of the differentiated planetary bodies in the inner Solar System is a topic of intense study and remains mostly unresolved. It is commonly thought to be the product of imperfect mixing of dust and gas during the early stages of the Solar System formation \citep[e.g.][]{Birck2004,AndreasenSharma2007,Trinquieretal2007,Brenneckaetal2013}. There have been several proposals put forth to explain the origin of this heterogeneity. The first of these is that it is a `cosmic chemical memory' that the solar nebula inherited from the natal molecular cloud \citep{Clayton1982,Dauphasetal2002}, which could have included the shell of a Wolf-Rayet star's wind bubble \citep{Dwarkadasetal2017}. 

The second group of proposals accounting for the heterogeneity calls for the late addition of external material from the interstellar medium. It has been suggested that fresh ejecta from asymptotic giant branch (AGB) stars or supernovae was added late into the solar nebula \citep{Trinquieretal2007}, or that there was a temporal change in the composition of infalling material from the molecular cloud, with early infalling material having compositions similar to the carbonaceous (C) meteorites group and later infalling material having compositions similar to the NC group \citep{Nanneetal2019}. 

The third group of proposals relates the isotopic heterogeneity to physical processes in the solar nebula itself. Thermal gradients in the protoplanetary disc could have selectively removed volatile elements, moderately volatile elements, and thermally unstable presolar silicates in the region closer to the Sun, creating a compositional gradient with distance from the Sun \citep[e.g.][]{Trinquieretal2009,Burkhardtetal2012,Eketal2020}. Furthermore, thermal processing could also modify the composition of infalling material from the molecular cloud \citep{Eketal2020}. This process has also been suggested to explain the temporal change observed in the composition of some elements such as Nd, which may be caused by the loss of SiC dust carriers between the time of formation of achondrite and later chondrite parent bodies within the inner Solar System \citep{Frossardetal2021}. This does not exclude an existing spatial heterogeneity of mass-independent variations, which are evident for other elements between achondrite groups whose parent bodies formed within 1.5 million years after the formation of the calcium-aluminium-rich inclusions (CAIs) \citep[e.g.][]{Luuetal2015}. Other than thermal gradients in the disc, the outward-then-inward movement of the water snow line throughout the lifetime of the gas disc has also been suggested as the mechanism that generates two generations of planetesimals with distinct compositions in the inner and outer Solar System \citep{Johansenetal2021,Lichtenbergetal2021}.

Recently, a new interpretation of the isotope data has been put forward. \citet{Schilleretal2018} suggested that the $\varepsilon$\element[][48]{Ca} isotopic heterogeneity among the Earth, Mars, and Vesta is correlated with the masses (and sizes) of these planetary bodies, and could instead reflect their different accretion timescales, that Vesta accreted earliest, followed by Mars, and finally the Earth. The authors suggest that the correlation is due to a change in the inner Solar System's composition and that the change could be brought about by the influx of pebbles from the outer Solar System. Assuming that the initial composition of the planetesimals in the inner Solar System is initially homogeneous and similar to those of the ureilites (a type of primitive and reduced achondrite with the lowest abundance of \element[][48]{Ca}), \citet{Schilleretal2018} proposed that the gradual accretion of CI carbonaceous chondrite-like material from the outer disc over the lifetime of the gas disc could account for the (different) isotopic compositions of Earth, Mars, and Vesta. This interpretation is at odds with the aforementioned works invoking radial heterogeneity in the disc to explain the observed isotopic differences of the major planetary bodies in the inner Solar System. The mechanism postulated by \citet{Schilleretal2018} formed the basis of further investigations by \citet{Johansenetal2021} who used semi-analytical simulations to study the feasibility of forming the terrestrial planets via the pebble accretion mechanism \citep[e.g.][]{OrmelKlahr2010,LambrechtsJohansen2012}.

The composition--distance correlation shown in Fig.~\ref{fig:correlation} is central to the discussion presented here. In this work we employ it to constrain plausible formation pathways of the planetary bodies in the inner Solar System that gave rise to the distinct isotopic compositions of Earth, Mars, and Vesta. Prior studies focusing on the formation of the terrestrial planets have demonstrated that it is useful to fuse cosmochemical observations with the outcomes of numerical simulations \citep[e.g.][]{NimmoAgnor2006,Nimmoetal2010,Rubieetal2015,Rubieetal2016,Brasseretal2017,Fischeretal2018,Wooetal2018,Zubeetal2019,Johansenetal2021,MahBrasser2021,Brennanetal2022}. These studies computed the bulk composition of the terrestrial planets in several dynamical models by tracking the regions in the disc from which the planets accreted and tying a particular region of the disc to a specific isotopic composition. The predicted compositions of the terrestrial planets from each model were then compared with cosmochemistry data to identify the best model(s) to describe the formation pathway of the planets. 

Works examining the classical model of planet formation \citep{Chambers2001} found that the feeding zones of the terrestrial planets (region of the protoplanetary disc where the planets sourced most of their building material) show a weak correlation with the semi-major axis \citep[e.g.][]{Raymondetal2004,Obrienetal2006,FischerCiesla2014,KaibCowan2015,Wooetal2018}. In this work the classical model is considered as the model that assumes that the gas giants are fully formed and that they remained close to their current orbits when the terrestrial planets were growing. This model is also known as the eccentric Jupiter and Saturn (EJS) model in several earlier works \citep[e.g.][]{Obrienetal2006,Raymondetal2009}. The correlation between planet feeding zone and orbital distance is then an expected outcome of this model because the terrestrial planets grew by the accretion of nearby solid material whose orbits were not strongly perturbed. Consequently, the isotopic differences of Earth and Mars can be reproduced if the material that these two planets accreted have different isotopic compositions. 

It is widely known, however, that the classical model has some shortcomings \citep[e.g.][]{Chambers2001,Raymondetal2009}; chief among them is its difficulty in producing a small Mars analogue with the correct mass at around 1.5~au. Subsequent proposals to overcome the shortcomings of the classical model can be roughly divided into two groups: the models that do not invoke the migration of the giant planets to achieve the desired mass and distribution of solids in the inner disc \citep[e.g.][]{Hansen2009,Izidoroetal2014,Izidoroetal2015} and the models that do \citep[e.g.][]{Walshetal2011,Clementetal2018}. The most recent model in the first group is the depleted disc model \citep{Izidoroetal2014,Izidoroetal2015}, which proposes a sharp drop in the inner disc's solid surface density with increasing heliocentric distance. In \citet{MahBrasser2021} we report that this model predicts a stronger correlation between terrestrial planet feeding zone and orbital distance compared to the EJS model due to the limited amount of material mixing in the disc. Models in the second group suggest a gas-driven inward-then-outward migration of Jupiter and Saturn \citep[Grand Tack model;][]{Walshetal2011}, a well-established model whose predictions we discuss below, or an early time for the dynamical instability (within 10~Myr after gas disc dispersal) among the giant planets in the outer Solar System \citep{Clementetal2018}, a model whose predictions have yet to be investigated.

Studies examining the Grand Tack model \citep{Walshetal2011} found that the feeding zones of the terrestrial planets are wide and strongly overlapping with no obvious correlation with the semi-major axis \citep{Brasseretal2017,Wooetal2018}. The main reason for this outcome is the migration of the gas giants. Jupiter and Saturn's excursions through the asteroid belt region excited the orbits of the material in the terrestrial planet region and caused that material to undergo mixing \citep{Carlsonetal2018}. If there was a difference in the composition of the material in the terrestrial planet region, it is expected that this difference would have been homogenised. The similar feeding zones of the terrestrial planets in the Grand Tack model thus imply that they should have similar compositions, in contrast with the isotope data. However, there is still a low but non-zero probability (<5\%) that Mars analogues can have feeding zones that are distinct to that of the Earth \citep{Brasseretal2017,Wooetal2018}. The terrestrial planet feeding zones computed from the EJS, depleted disc, and Grand Tack models suggest that the formation pathway of the terrestrial planets is more likely to follow the EJS and depleted disc models; in other words, the planets accreted the majority of their building blocks locally within their feeding zone with limited mixing of material in the disc. This is backed up by very recent high-resolution simulations run on graphics cards \citep{Wooetal2021}.

If Earth's and Mars' isotopic differences were indeed the consequences of dominantly local accretion, then the trend in Fig.~\ref{fig:correlation} would suggest that Vesta's feeding zone is isolated from that of Earth and Mars. However, it is not obvious, a priori, that Vesta formed where it is today: by looking only at the isotopic data, we cannot rule out the possibility that Vesta is a mixture of material from the terrestrial region and material from further out in the disc. Earlier works suggested that it could have originated from the region near Venus and the Earth, and that it was subsequently scattered outwards into
the asteroid belt by planetary embryos, defined here as planetary bodies of diameter $D\sim\mathcal{O}(10^3~{\rm km})$, in the terrestrial planet region \citep{Bottkeetal2006,Mastrobuono-BattistiPerets2017,RaymondIzidoro2017}. Alternatively, it could have been scattered inwards to its current orbit from the outer Solar System when the gas giants were growing \citep{Izidoroetal2016}. Investigating the dynamical outcomes of terrestrial planet formation models with a focus on the asteroid belt region while taking into account the constraints from available isotopic data could therefore provide additional insights to the problem at hand.

However, the condition of local accretion alone is insufficient to explain the isotopic differences between these three planetary bodies. For example, if the initial composition of the inner disc were homogeneous, that is, if the nucleosynthetic isotopes were equally distributed among all the solid material, then the resultant composition of all the three planetary bodies would be identical irrespective of whether mixing occurred. For this reason, another condition is required in addition to local accretion to fully explain the reported correlation: the distribution of nucleosynthetic isotopes in the inner disc is heterogeneous. Several works \citep{Yamakawaetal2010,Mezgeretal2020,Spitzeretal2020} have suggested that nucleosynthetic anomalies in the inner Solar System were distributed along a gradient and that this isotopic gradient was established at the time before the formation of the planetesimals that were the precursors of the terrestrial planets and the asteroids.

In this work we look into the formation locations of asteroids Vesta, Hungaria, and Hebe within the framework of various dynamical terrestrial planet formation models, and determine which models support local formation of asteroids such as Vesta. To do this, we tap into our available database containing {\it N}-body simulation outputs for the Grand Tack, EJS, and depleted disc models. We consider the EJS model as a comparison rather than a viable hypothesis. These simulations were initially performed to study the formation of the terrestrial planets and the results can be found in \citet{Brasseretal2016}, \citet{Wooetal2018,Wooetal2021}, and \citet{MahBrasser2021}. The data analysis is followed by a discussion of two important additional conditions, the presence and preservation of an isotopic gradient in the inner disc, which are required to account for the observed trend between isotopic anomalies and distance from the Sun in the inner Solar System. 

\section{Methods}
\begin{table*}
\centering
    \caption{Summary of our numerical simulation database.}
    \label{table:summary}      
    \begin{tabular}{l l l c c c c}     
    \hline\hline       
    Model & Resolution & References & No. of simulations & Hungaria & Vesta & Hebe\\ 
    \hline                    
    Grand Tack (GT)    & Low  & 1 & 336 & 3920 & 2233 & 2119\\  
    EJS                & Low  & 2     & 128 & 2072 & 5294 & 5536\\
                       & High & 3     & 18  & 141  & 58   & 53\\
    Depleted disc (DD) & Low  & 4 & 144 & 5987 & 8623 & 10238\\
    \hline                  
    \end{tabular}
    \tablebib{(1) \citet{Brasseretal2016}; (2) \citet{Wooetal2018}; (3) \citet{Wooetal2021}; (4) \citet{MahBrasser2021}.}
\end{table*}

\subsection{Numerical simulation database}
Here we provide a brief description of the initial conditions of the numerical simulations in our database.
\subsubsection{Grand Tack model simulations}
The simulations for the Grand Tack model \citep[from][]{Brasseretal2016} start with Jupiter and Saturn migrating inwards through the asteroid belt in the first 0.1~Myr after the beginning of the Solar System, as given by the formation of the CAIs. When Jupiter reaches 1.5~au, it reverses its migration direction and ushers Saturn along until they reach their proposed semi-major axes at $\sim 5.4~{\rm au}$ and $\sim 7.5~{\rm au}$, respectively \citep{Morbidellietal2007}, before the late giant planet instability. 

Within the orbits of Jupiter and Saturn, a solid disc composed of embryos and planetesimals having nearly circular and coplanar orbits was placed between 0.7 and 3~au. The surface density of embryos and planetesimals follows $\Sigma_{\rm solid} \propto r^{-3/2}$. The initial masses of the embryos are either identical throughout the disc \citep{JacobsonMorbidelli2014}, with the total mass ratio of embryos to planetesimals being 1:1, 4:1, or 8:1 (the masses of the individual embryos are 0.025, 0.05, and 0.08 Earth mass, respectively), or are computed using a semi-analytic oligarchic growth model \citep{Chambers2006} where the mass of the embryos and the spacing between them (5, 7, or 10 mutual Hill radii) are dependent on the their isolation mass \citep{KokuboIda1998}. The planetesimals in the oligarchic growth model simulations have a mass of $M=10^{-3}$ Earth mass and diameter of 1430~km. The initial density of the embryos and planetesimals are assumed to be $3\,000~{\rm kg~m^{-3}}$. 

The simulations were run for a total of 150~Myr using a time step of 0.02~yr with the presence of a gas disc in the first 5~Myr. The gas disc model adopted was based on that of \citet{Bitschetal2015}, which features a higher surface gas density than that employed by \citet{Walshetal2011} in their first Grand Tack simulations. The initial gas surface density profile $\Sigma_{\rm gas} \propto r^{-1/2}$ and the temperature profile $T(r) \propto r^{-6/7}$. The gas surface density and temperature profiles both decay very rapidly in the first Myr followed by a slower decay until $t = 5~{\rm Myr}$ in the simulation, after which the disc is artificially photoevaporated over the next 100~kyr.

\subsubsection{Classical (EJS) model simulations}
The initial conditions for the EJS model simulations of \citet{Wooetal2018} are similar to those for the Grand Tack model, except that Jupiter and Saturn stayed on their current orbits throughout the simulations and the gas disc was not included. In that study the authors only investigated the case where the embryos were assumed to have undergone oligarchic growth \citep{Chambers2006} and are spaced 5 or 10 mutual Hill radii apart.

A more recent study of the EJS model \citep{Wooetal2021} employed the GENGA {\it N}-body code, which runs on GPUs \citep{GrimmStadel2014}. Similar to \citet{Wooetal2018}, the initial solid surface density of the disc follows the minimum mass solar nebula \citep[MMSN;][]{Hayashi1981}. However, the initial solid disc of \citet{Wooetal2021} has a much higher resolution than \citet{Wooetal2018}; it is assumed to consist of a bimodal population of equal mass planetesimals $(R = 350~{\rm km}~{\rm or}~800~{\rm km})$ with nearly circular and coplanar orbits $(e < 0.01~{\rm and}~ i < 0.5\degr)$ distributed between 0.5 and 3~au, all of which are fully self-gravitating. The simulations are performed for 150~Myr.

In the \citet{Wooetal2021} simulations a gas disc is included with a surface density profile of 
\begin{equation}
    \Sigma_{\rm gas}(r,t) = \Sigma_{\rm gas,0}(r/1~{\rm au})^{-p}\exp(-t/\tau_{\rm decay})\,,
\end{equation}
where $\Sigma_{\rm gas,0}$ is the initial gas surface density at 1~au and $\tau_{\rm decay}$ is the timescale for gas decay. At $t=0$, the initial gas surface density $\Sigma_{\rm gas,0}$ is assumed to be $2\,000~{\rm g~cm^{-2}}$ and $p=1$, which corresponds to the nominal disc from \citet{Morishimaetal2010}. The gas disc decay timescale $\tau_{\rm decay}$ was chosen to be either 1 or 2~Myr. The effects of gas drag and Type I migration were also included.

\subsubsection{Depleted disc model simulations}
For the depleted disc model, the initial conditions from \citet{MahBrasser2021} are also broadly similar to the EJS model simulations of \citet{Wooetal2018}, but with several differences: (1) the inner edge of the solid disc was extended further inwards to 0.5~au; (2) the solid surface density beyond 1, 1.25, or 1.5~au was depleted by different amounts (50\%, 75\%, and 95\%) with respect to the MMSN; and (3) the gas disc model was adopted from \citet{Idaetal2016}, with surface density and temperature profiles following $\Sigma_{\rm gas} \propto r^{-3/5}$ and $T(r) \propto r^{-9/10}$ in the inner regions, and $\Sigma_{\rm gas} \propto r^{-15/14}$ and $T(r) \propto r^{-3/7}$ in the outer regions. After 5~Myr, the gas disc is assumed to have photoevaporated away completely and the simulations were continued for another 150~Myr without the gas disc. The time step used in the simulations is 0.01~yr.

In our simulations of the three dynamical models mentioned above, we employed different prescriptions for the gas disc. Although the Sun's primordial gas disc is unconstrained, previous works \citep[e.g.][]{Morishimaetal2010,WalshLevison2019} have shown that variations in gas disc parameters (e.g. gas disc decay timescale $\tau$) do not produce drastically different simulation outcomes in terms of the dynamics of the final planetary system. We therefore do not expect that the choice of $\tau$ (whether 1~Myr or 2~Myr) to make a huge difference in what is being analysed for the purpose of this paper.

Our simulations lack the necessary resolution to study the formation of objects the size of Hebe $(D \approx 200~{\rm km})$ or Vesta $(D \approx 500~{\rm km})$. The planetesimals in most of the simulations are $10^{-3}$ Earth mass with diameter $D \approx 1400~{\rm km}$, which is already much larger than the size of the asteroids of interest here. Even for the high-resolution GPU simulations, the minimum planetesimal diameter is 700~km, which is slightly larger than Vesta. Furthermore, with the exception of the high-resolution simulations \citep{Wooetal2021}, all the planetesimals in the simulations are not allowed to interact gravitationally with each other and thus we cannot retrieve any information on their accretion histories from the simulations. Our focus in this work is therefore on the orbital evolution of planetesimals that end up near the modern orbits of Vesta, Hebe, and Hungaria.

\subsection{Data analysis}
In total, we analysed 336 simulations from \citet{Brasseretal2016} for the Grand Tack model, 146 simulations from \citet{Wooetal2018,Wooetal2021} for the EJS model, and 144 simulations from \citet{MahBrasser2021} for the depleted disc model. 

We defined all the planetary bodies in our
simulations, regardless of their final mass, as Hungaria, Vesta, and Hebe analogues if their final semi-major axes $a_{\rm f}$ fulfil the following criteria:
\begin{itemize}
    \item Hungaria: 1.7~au < $a_{\rm f}$ < 2.1~au;
    \item Vesta: 2.1~au < $a_{\rm f}$ < 2.5~au;
    \item Hebe: 2.2~au < $a_{\rm f}$ < 2.6~au.
\end{itemize}

We find that for the Grand Tack model, planetary embryos constitute 2\% of the total number of asteroid analogues that end up in the disc regions of interest defined above, while the fraction is 0.9\% for the EJS model and 0.3\% for the depleted disc model. The fraction of planetary embryos that end up in the region beyond 1.7~au is highest for the Grand Tack model because Jupiter and Saturn's migration reshuffles the orbits of the solid material in the disc \citep{Carlsonetal2018}. We included all objects (planetary embryos and planetesimals) in the analysis of this work.

Finally, we track the initial semi-major axis of the asteroid analogues. We filtered out objects with an initial semi-major axis greater than 3~au because that is defined as the outer edge of the solid disc in our simulations. We summarise the total number of asteroid analogues found in the simulations in Table~\ref{table:summary}. There are on average many more asteroid analogues in the depleted disc model compared to both the Grand Tack and the EJS models because of the depletion in mass and the lack of massive planetary embryos in the region beyond the orbit of Mars so that no efficient scattering and mixing can occur. This in turn limits the degree of mixing among the solids and thus the majority of the planetesimals in this region tend to remain where they are throughout the simulations.

\section{Results and discussion}
\subsection{Formation region of asteroids}
\begin{figure}
\centering
   \includegraphics{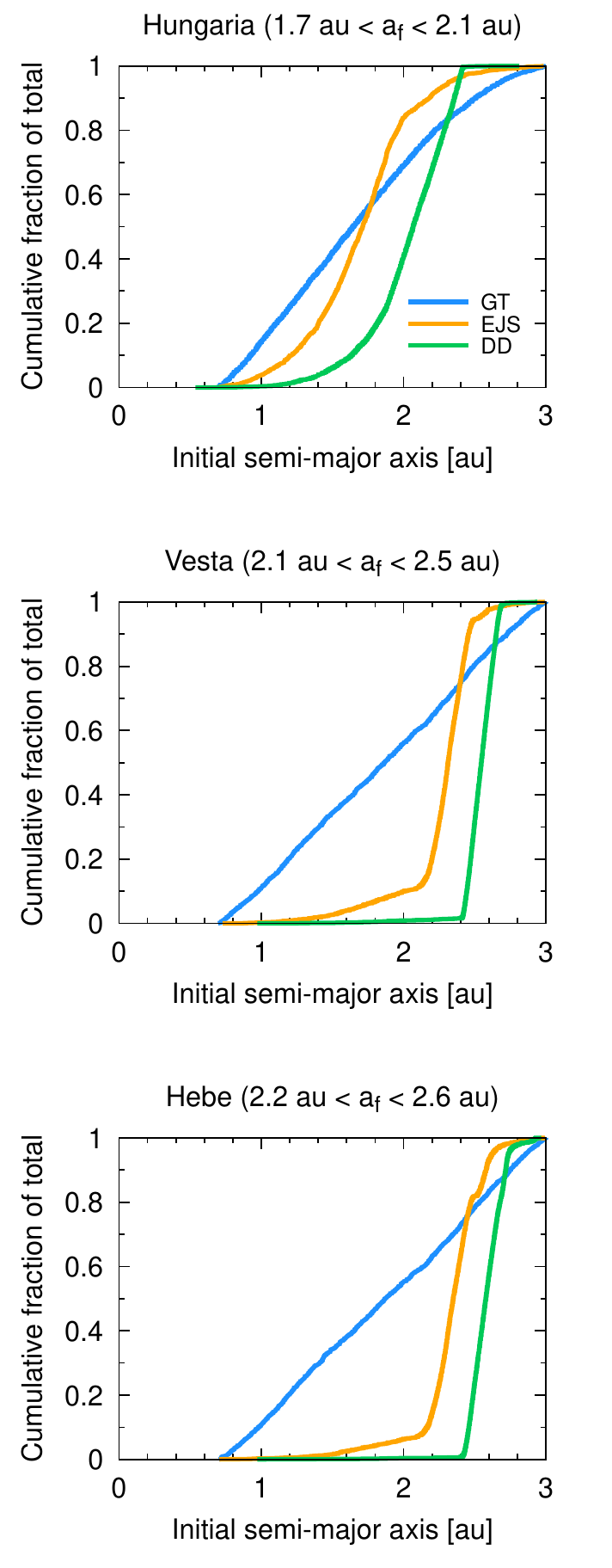}
   \caption{Cumulative distributions of the initial semi-major axis of Hungaria, Vesta, and Hebe analogues in the Grand Tack (GT), EJS, and depleted disc (DD) models. The asteroids have equal probability of originating from anywhere in the disc for the Grand Tack model, while their formation locations are more restricted in the EJS model and the depleted disc model.}
   \label{fig:asteroid_init_a_cumulative}
\end{figure}
We present the results for the formation location of asteroid analogues Hungaria, Vesta, and Hebe obtained from all of the simulations of the Grand Tack, EJS, and depleted disc models in Fig.~\ref{fig:asteroid_init_a_cumulative}. 

In the Grand Tack model all the analogues of asteroids Hungaria, Vesta, and Hebe have initial semi-major axes ranging from 0.7 to 3.0~au with roughly equal probability (the cumulative distribution is almost a straight line), corresponding to the inner and outer edge of the solid disc (top panel of Fig.~\ref{fig:asteroid_init_a_cumulative}). The probability of Hungaria analogues coming from beyond 1.7~au is 47.0\%; the probability for Vesta to form beyond 2.1~au is 39.7\%; there is a 36.8\% chance that Hebe originated from beyond 2.2~au. The near-linear trend of the cumulative distribution can be attributed to the complete mixing of solid material in the disc which is brought about by the gas-driven migration of Jupiter and Saturn \citep{Carlsonetal2018}. 

In the EJS and the depleted disc models, the initial semi-major axes of the asteroid analogues are severely restricted. For the EJS model, 51.1\% of the Hungaria analogues come from beyond 1.7~au, 89.1\% of the Vesta analogues come from beyond 2.1~au, and 85.6\% of the Hebe analogues come from beyond 2.2~au. The probabilities for the same analogues in the depleted disc model are 86.8\%, 99.0\%, and 99.6\%, the highest of the three dynamical models studied here. The higher probabilities of asteroid analogues in these two models originating near or beyond their current orbits are in contrast to the outcome from the Grand Tack model. This can be attributed to the lower extent of material mixing in the solid discs as Jupiter and Saturn did not migrate into the inner Solar System when the gas disc was present. In the case of the depleted disc model, the small amount of solid material beyond the orbit of Mars further reduces the degree of mixing in the disc. 

\begin{figure}
\centering
   \includegraphics{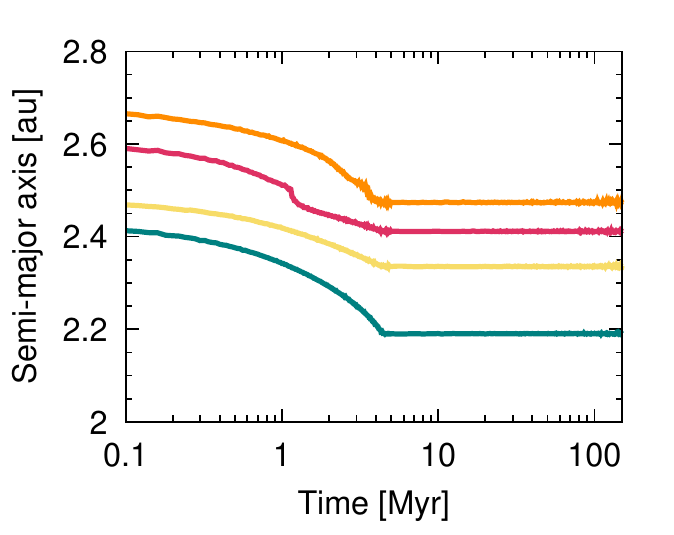}
   \caption{Evolution of the semi-major axes of four sample Vesta analogues taken from one of our depleted disc model simulations. The Vesta analogues in this particular run were initially located in the region between 2.4~au to about 2.66~au in the disc, and then migrated inwards during the first 5~Myr of the simulation under the effects of gas drag and orbital eccentricity damping before settling at their final orbits.}
   \label{fig:vesta_init_a_example}
\end{figure}

One would thus expect that Hungaria, Vesta, and Hebe should form very close to their current orbits in the depleted disc model because of the very low level of mixing in the disc. We found, however, that this is not the case and that the asteroid analogues predominantly originate from wider orbital distances instead. The reason for this can be traced to the effects of the gas disc in the first 5~Myr of the simulations; the combined effects of gas drag and orbital eccentricity damping exerted by the gas disc on the planetesimals caused them to migrate inwards during the first 5~Myr. We show in Fig.~\ref{fig:vesta_init_a_example} how the semi-major axes of Vesta analogues in one of our depleted disc model simulations change with time.

For all the three dynamical models, the Hungaria analogues have a wider cumulative distribution for their initial semi-major axis compared to that of the Vesta and Hebe analogues. This is the result of dynamical scattering events with larger embryos that are present in the Hungarian region of our simulations, but not in the main belt.

\subsection{Vesta as a native of the asteroid belt}
For each dynamical model, we further scoured our database to find planetary systems that most closely resemble the current Solar System, and then determine the formation location of the Vesta analogues in these systems. These best Solar System analogues must have at least one analogue each for Venus, Earth, and Mars \citep[based on the mass--distance criteria given in][]{Brasseretal2016} and three out of four statistics (AMD, mass concentration parameter, mass fraction of the largest planet, mean orbital spacing parameter) introduced by \citet{Chambers2001} falling within $2\sigma$ of the current Solar System value. We find one system each for the Grand Tack and depleted disc models that fulfil these criteria. For the EJS model, we relaxed the Chambers statistics criterion to two (instead of three) out of four statistics because most of the final planetary systems fulfil only one of the four statistics. We find that two systems (one each from the low- and high-resolution simulations) fulfil this criterion.

\begin{figure}
\centering
   \includegraphics{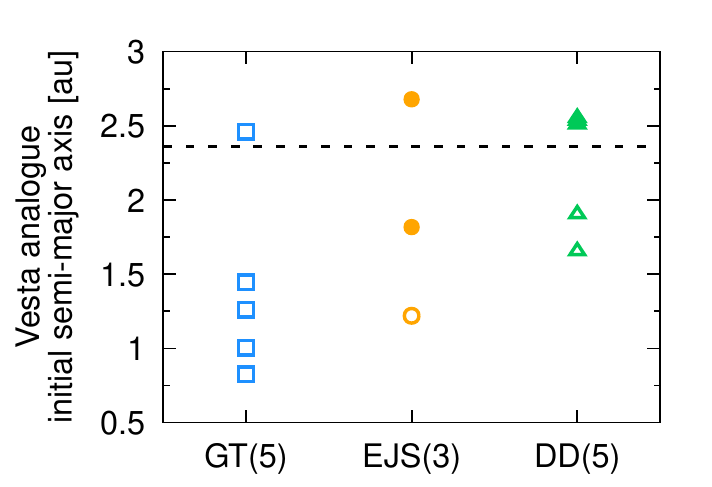}
   \caption{Initial semi-major axes of all the Vesta analogues from the best outcomes of the different dynamical models. Vesta analogues from high-resolution EJS model simulations are indicated with filled circles. Numbers in parentheses are the total number of Vesta analogues. The black dashed line shows the current semi-major axis of Vesta (2.36~au).}
   \label{fig:vesta_best_allmodels}
\end{figure}

In Fig.~\ref{fig:vesta_best_allmodels} we plot the initial semi-major axis of the Vesta analogues from our best Solar System analogues. The best Grand Tack model output has five Vesta analogues, four of which came from within the orbit of Mars and only one originated from the asteroid belt region. For the EJS model, the best output from the low-resolution simulations contains only one Vesta analogue which came from the terrestrial planet region, while the Vesta analogues from the best output from the high-resolution simulations formed in the region beyond the orbit of Mars. For the depleted disc model, all five of the Vesta analogues from the best output have initial semi-major axes larger than 1.5~au. Except for our best Grand Tack model output, the outcomes from our best EJS and depleted disc model simulations are consistent with the overall trend we presented earlier in Fig.~\ref{fig:asteroid_init_a_cumulative}: it is more likely for a Vesta analogue to form close to its current orbit in models where the degree of material mixing in the solid disc is low. 

\begin{figure}
\centering
   \resizebox{\hsize}{!}{\includegraphics{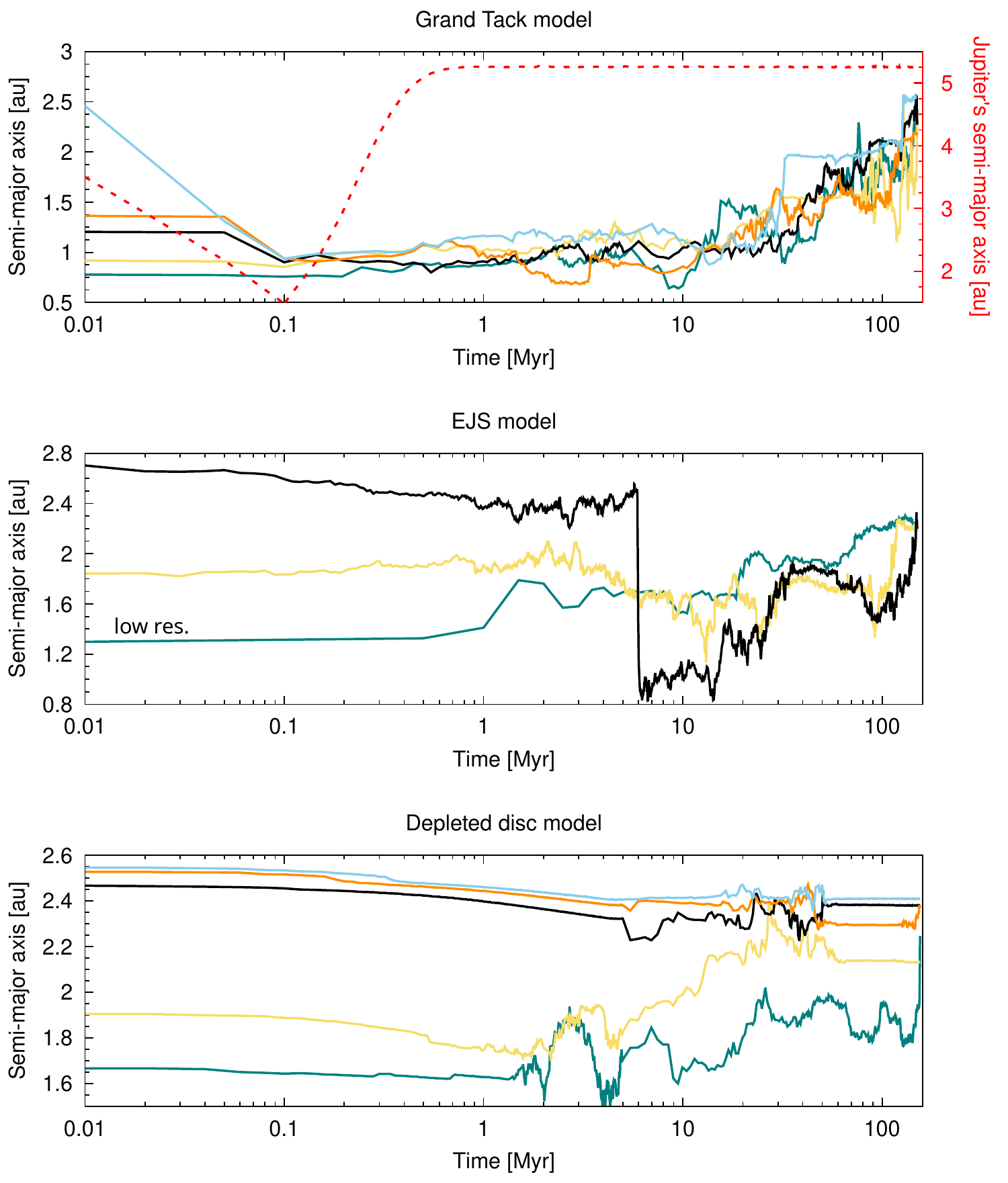}}
   \caption{Evolution of the semi-major axes over time for all the Vesta analogues from each dynamical model's best Solar System analogue. For the Grand Tack model (top panel) we also plot the evolution of Jupiter's semi-major axis with red dashed line.}
   \label{fig:vesta_best_a_evo}
\end{figure}

The time evolution of the semi-major axes of each of the Vesta analogues in the best model outputs are depicted in Fig.~\ref{fig:vesta_best_a_evo}. In the best Grand Tack model simulation, Jupiter and Saturn's gas-driven migration during the first 0.1~Myr of the simulation pushed the Vesta analogues with initial orbits further than 1.2~au to within 1~au (top panel of Fig.~\ref{fig:vesta_best_a_evo}). The Vesta analogues were then scattered around in the disc due to the reshuffling of the orbits of the solid material in the same region induced by the movement of the gas giants before being pushed to wider orbital distances by the growing planetary embryos in the terrestrial planet region. 

In the best low-resolution EJS model simulation, the sole Vesta analogue originating from 1.3~au also displays an evolution similar to the Vesta analogues from the Grand Tack simulation, that is, scattered out to its final orbit by planetary embryos located in the terrestrial planet region. On the other hand, Vesta analogues from the best high-resolution simulation originate from beyond 1.5~au, and thus have different evolution pathways. They spent the first few million years near their initial orbits and then at 4~Myr were deposited into the terrestrial planet region by the sweeping of the $\nu_5$ secular resonance \citep{Wooetal2021} before being scattered back to the asteroid belt.

In the best depleted disc model simulation, the Vesta analogues located initially beyond 2.4~au migrated inwards initially as they exchanged angular momentum with the surrounding gas. After the gas disc dispersed at $t = 5~{\rm Myr}$, they were scattered about but remained close to their initial orbits because of the low mass and density of solids in that region of the disc. In contrast, Vesta analogues with initial semi-major axes within 2~au began to be scattered around earlier before gas disc dispersal, due to the presence of planetary embryos in the same region that were actively accreting planetesimals. After $t = 5~{\rm Myr}$, the planetary embryos began to scatter each other and clear the terrestrial planet region of planetesimals, resulting in the continuous outward migration of the two Vesta analogues to their final orbital distances.

As none of our simulations include solid material exterior to the orbit of Jupiter and Saturn, unlike the simulations of \citet{RaymondIzidoro2017}, we are therefore unable to fully evaluate the possibility of any of the asteroid analogues of interest in this work originating from the outer Solar System. However, an origin of Vesta in the outer Solar System is rather unlikely from the cosmochemical point of view given the very distinct isotopic compositions of the HED meteorites compared to the carbonaceous group of meteorites \citep[e.g.][]{Warren2011}. If Vesta had formed in the outer Solar System, its isotopic composition would be expected to be similar to the carbonaceous chondrites. By the same reasoning, Vesta's distinct isotopic composition compared to the terrestrial planets also suggests that it probably could not have come from within 1.5~au. Instead, Vesta's isotopic similarity to several other achondritic meteorite groups such as the brachinites and angrites in their oxygen isotopic compositions \citep{ClaytonMayeda1996} suggests an origin in a location that is different from both the terrestrial planets (inner Solar System) and the carbonaceous chondrites (outer Solar System). We therefore argue that Vesta is likely to have formed locally in the asteroid belt and is a `native' there. The above reasoning requires a heterogeneous disc composition because a homogeneous composition would not result in distinct isotopic compositions of the Earth and Mars.

Our simulation results for the dynamical models not involving the gas-driven migration of the gas giants (i.e. EJS model and depleted disc model) show that Vesta has a higher probability of forming in the asteroid belt, in support of the previous cosmochemistry argument for Vesta having formed in a different region from the terrestrial planets and the carbonaceous chondrites. However, there is also a reasonable fraction of Vesta, Hebe, and Hungaria analogues that originate from the asteroid belt region in the Grand Tack model. When taking into account the terrestrial planet feeding zone trends for the Grand Tack, EJS and depleted disc models reported earlier in \citet{Brasseretal2017}, \citet{Wooetal2018}, and \citet{MahBrasser2021}, our results suggest that dynamical models in which there is no strong mixing of solid material in the disc are more likely to be consistent with the distinct isotopic compositions of the Earth, Mars, and Vesta. This conclusion hinges on whether the asteroids Vesta and Hebe (possibly) and the Hungaria group are representative of all the asteroids or not. At present, Vesta is the only asteroid that can be linked with a high confidence level to meteorites from the currently available pool of samples. Our conclusion here would be different if we sampled a large number of asteroids and found that they have different formation locations.

\subsection{Explaining the distance-composition correlation}
The condition of local accretion (or local formation) alone is, however, insufficient to fully explain the isotopic data trends (Fig.~\ref{fig:correlation}) discussed in the Introduction, where we mentioned that two other conditions need to be fulfilled simultaneously.

The first condition is the presence of an isotopic gradient in Ca, Cr, Ti, and possibly Mo \citep{Regelousetal2008,Trinquieretal2009,Yamakawaetal2010,Schilleretal2018,Spitzeretal2020} in the inner Solar System such that when the planetary seeds of Earth, Mars, and Vesta accreted their building blocks, their final isotopic compositions would naturally reflect the nature of the distribution of the isotopes in the solid disc. It thus goes without saying that the distribution of isotopes in the disc should be heterogeneous. A homogeneous distribution of isotopes would not result in any differences in the isotopic compositions of Earth, Mars, and Vesta regardless of whether these planetary bodies accreted locally \citep{Wooetal2018}.

The isotopic gradient should exist before the formation of the planetesimals because subsequent collisional growth to form larger objects will only produce planetary bodies with compositions of refractory elements that are the average of the planetesimals they accreted. The gradient was likely to have been established by $\sim 4566~{\rm Ma}$ (that is, within 1-2~Myr after CAIs) based on the available chronology data for the following events:
\begin{itemize}
    \item Ages of CAIs: 4567-4568 Ma \citep[e.g.][]{Amelinetal2010,BouvierWadhwa2010,Connellyetal2012};
    \item Formation of iron meteorite and angrite parent bodies: >4566 Ma \citep[e.g.][]{Connellyetal2008,Kleineetal2012,Hansetal2013,Kruijeretal2014,Kruijeretal2017,SugiuraFujiya2014,KruijerKleine2019};
    \item Ages of most chondrules: $\sim 4566~{\rm Ma}$ \citep[e.g.][]{Luuetal2015};
    \item Accretion of Vesta: $0.8 \pm 0.3~{\rm Myr}~(\sim ~4566~{\rm Ma})$ to maximum 1.5~Myr after CAIs \citep[e.g.][]{Neumannetal2014,SugiuraFujiya2014} and subsequent differentiation and magmatism recorded over the next 50~Myr by eucrite crystallisation ages \citep{Bouvieretal2015};
    \item Accretion of Mars: $\sim 44\%$ of its present size within a strict lower limit of 1.2~Myr after Solar System formation $(\sim 4566~{\rm Ma})$ \citep[e.g.][]{DauphasPourmand2011,TangDauphas2014}; 
    \item Protoplanetary disc dissipation: $\sim 4563~{\rm Ma}$ \citep{Wangetal2017}.
\end{itemize}

\begin{figure*}
\centering
   \resizebox{0.9\hsize}{!}{\includegraphics{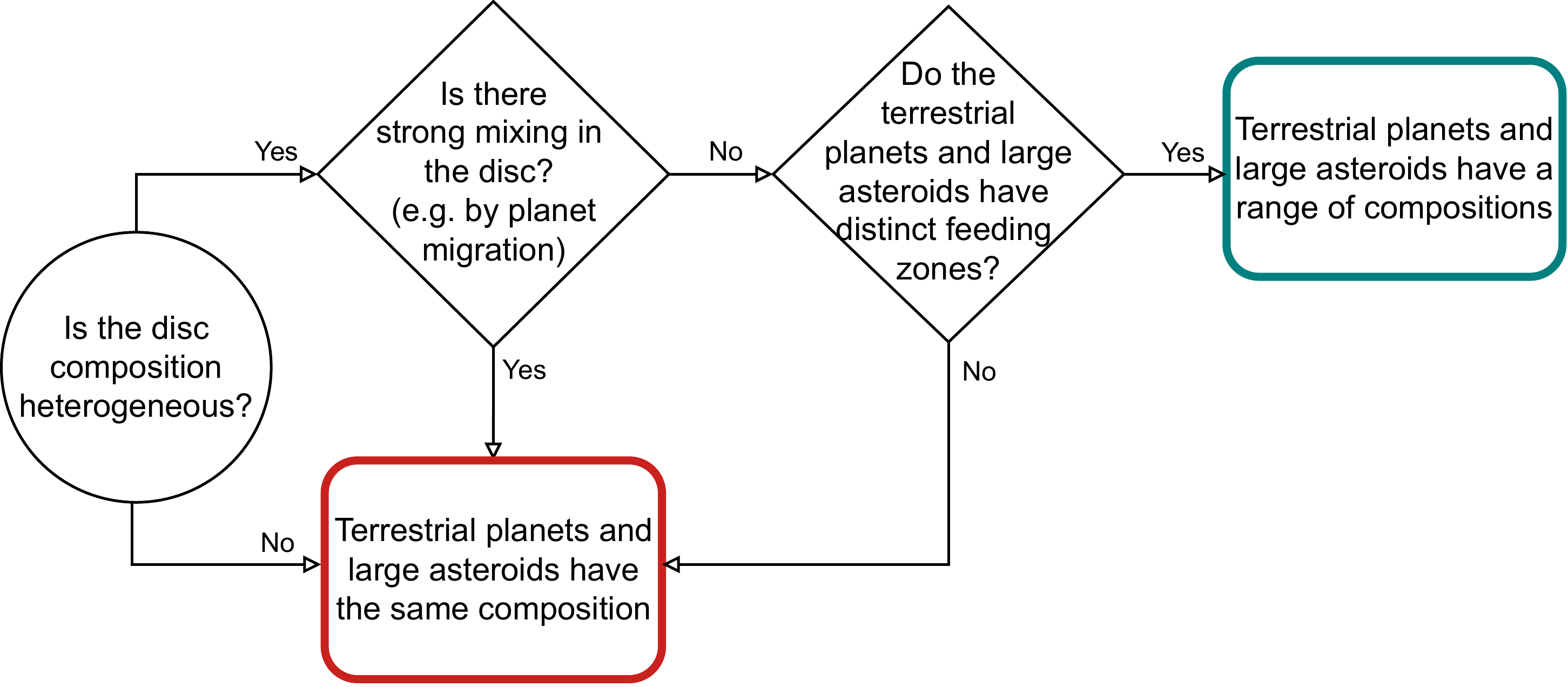}}
   \caption{Flow chart summarising the three conditions required to produce the distinct and distance-correlated isotopic compositions of the Earth, Mars, and Vesta. The most important condition is the heterogeneous distribution of isotopes along a gradient in the inner Solar System. If this condition is not fulfilled, then the planetary bodies will have the same isotopic composition regardless of whether there was mixing in the disc or whether the feeding zones are distinct.}
   \label{fig:flowchart}
\end{figure*}

The second condition is the preservation of the isotopic gradient during the formation of the planetesimals, the terrestrial planets and Vesta. If the gradient was disturbed or homogenised, for example by the migration of Jupiter and Saturn, then the planetary bodies that grew from the mergers of planetesimals would not display distinct isotopic compositions, as was observed in the data, despite having accreted their building blocks locally. In Fig.~\ref{fig:flowchart} we provide a flowchart as a comprehensive explanation of the trend found in Fig.~\ref{fig:correlation}.

\section{Conclusions}
The apparent correlation between semi-major axis and some mass-independent isotopic anomalies ($\varepsilon$\element[][48]{Ca}, $\varepsilon$\element[][50]{Ti}, $\varepsilon$\element[][54]{Cr}, $\varepsilon$\element[][92]{Mo}) for the Earth, Mars, and Vesta can be used to constrain plausible dynamical mechanisms for planet formation in the inner Solar System. Here we presented a study combining dynamical modelling and cosmochemical arguments to explain the observed correlation. We find that dynamical models where the materials in the inner Solar System did not undergo strong mixing are more likely to produce the terrestrial planets with distinct feeding zones and asteroids with restricted formation location. This result, coupled with the presence of an isotopic gradient before the formation of the planetesimals and the preservation of the gradient throughout the formation of the planetary bodies, will result in the distinct isotopic compositions of the Earth, Mars, and Vesta shown in the cosmochemistry data. Our conclusion relies on Vesta to be representative of all asteroids. In the Grand Tack model, where the gas-driven migration of Jupiter and Saturn is called upon to truncate the inner disc, the mixing of solid material in the disc would dilute any isotopic heterogeneities and result in the Earth, Mars, and Vesta all having the same isotopic composition, which is inconsistent with the data. We suggest therefore that it was rather unlikely that the gas giants migrated to the inner Solar System when the gas disc was present. The migration of the giant planets could have happened at a later time \citep{Clementetal2018}, and its effect on the degree of mixing in the inner disc deserves detailed investigation in the future. Finally, the dynamical evolution of Vesta and its unique isotopic composition provide an additional powerful constraint for models of terrestrial planet formation.

\begin{acknowledgements}
      J.M. acknowledges the support of the DFG priority program SPP 1992 ``Exploring the Diversity of Extrasolar Planets'' (BI 1880/3-1). R.B. and S.J.M. thank the Research Centre for Astronomy and Earth Sciences (Budapest, Hungary) for support of the Origins Research Institute (ORI). We thank Matt Clement for his careful review and constructive comments that helped improve this manuscript.
\end{acknowledgements}

\bibliographystyle{aa} 
\bibliography{main} 


\end{document}